\documentclass[iop]{emulateapj} \usepackage{apjfonts}

\usepackage{amsmath,amssymb,amsfonts} \usepackage{mathrsfs}
\usepackage{graphicx} \usepackage{subfigure}
\newcommand{\be}{\begin{equation}} \newcommand{\ee}{\end{equation}}

 \newcommand{\lp}{\left(}
\newcommand{\rp}{\right)} \newcommand{\bb}{\begin{bmatrix}}
\newcommand{\eb}{\end{bmatrix}} \DeclareMathOperator{\Tr}{Tr}

\begin{document}

\title{An Efficient Approximation to the Likelihood for
Gravitational\\ Wave Stochastic Background Detection Using Pulsar
Timing Data}

\author{J. A. Ellis\altaffilmark{1}, X. Siemens\altaffilmark{1}, and
R. van Haasteren\altaffilmark{2}}

\altaffiltext{1}{Center for Gravitation, Cosmology and Astrophysics,
University of Wisconsin Milwaukee, Milwaukee WI, 53211}

\altaffiltext{2}{Max-Planck-Institut f\"{u}r Gravitationphysik
(Albert-Einstein-Institut), D-30167 Hanover, Germany}

\begin{abstract} 
Direct detection of gravitational waves by pulsar
timing arrays will become feasible over the next few years. In the low
frequency regime ($10^{-7}$ Hz -- $10^{-9}$ Hz), we expect that a
superposition of gravitational waves from many sources will manifest
itself as an isotropic stochastic gravitational wave background.
Currently, a number of techniques exist to detect such a signal;
however, many detection methods are computationally challenging. Here
we introduce an approximation to the full likelihood function for a
pulsar timing array that results in computational savings proportional
to the square of the number of pulsars in the array. Through a series
of simulations we show that the approximate likelihood function
reproduces results obtained from the full likelihood function. We
further show, both analytically and through simulations, that, on
average, this approximate likelihood function gives unbiased parameter
estimates for \emph{astrophysically} \emph{realistic} stochastic
background amplitudes. 
\end{abstract}

\maketitle

\section{Introduction}

Gravitational waves (GWs) will very likely be detected in the next few
years. Pulsar timing arrays (PTAs)~\citep{haa+10} as well as
ground-based interferometers such as Advanced
LIGO~\citep{Waldman:2011vg} are expected to make the first direct GW
detection on a similar time-scale, though they are sensitive to
different and complementary regions of the GW spectrum. Ground-based
instruments are most sensitive around 100~Hz, and the most promising
source at those frequencies are binaries of compact objects such as
neutron stars and black holes (up to a few tens of solar masses).
Pulsar timing arrays are most sensitive around $10^{-9}$~Hz, and the
most promising source at those frequencies are super-massive binary
black holes (SMBBHs) that coalesce when galaxies merge.

All the SMBBH mergers that have taken place throughout the history of
our universe produce a stochastic background of gravitational
waves~\citep{lb01,jb03,wl03,vhm03,ein+04,svc08,s13,McWilliams:2012jj},
as well as individual periodic signals that may be detectable as above
the confusion noise~\citep{svv09,sv10,rs11,rwh+12,Mingarelli:2012hh},
and bursts~\citep{vl10,Cordes:2012zz}. A number of techniques have
been implemented to search pulsar timing data for the stochastic
background~\citep{det79,srt+90,l02,jhl+05,jhs+06,abc+09,hlm+09,vlm+09,ych+11,vhj+11,Cordes:2011vg,dfg+12},
as well as periodic signals~\citep{jll+04,yhj+10,cc10,lwk+11,
ejm12,bs12,esc12,pbs+12}, and bursts~\citep{fl10}.

For stochastic background searches, evaluations of the full likelihood
are computationally challenging. PTAs are currently timing up to a few
tens of pulsars, with several thousand points each. In addition, the
likelihood function depends not only on the relatively small number of
parameters that characterize GW stochastic background, but also on
several intrinsic red and white noise parameters for each pulsar. A
number of techniques have already been introduced to reduce the
computational burden of such
searches~\citep{vh12,Lentati:2012xb,Taylor:2012vx}, and we will
discuss these results later in the paper.

Although the stochastic background produces random changes in the
times-of-arrival (TOAs) of an individual pulsar, the cross-correlation
of its effects on two pulsars only depends on the angular separation
between pulsars~\citep{hd83}. In this paper we introduce an efficient
approximation to the likelihood by using an expansion to first order
in the amplitude of the cross-correlation terms introduced
by~\cite{abc+09}. This technique has already used to analyze the first
International Pulsar Timing Array Mock Data Challenge~\citep{esc12b}.
The approximation affords us a computational savings quadratic in the
number of pulsars in the pulsar timing array, a factor of a one to
three orders of magnitude, depending on the size of the PTA.

This paper is organized as follows. In Section~\ref{sec:timingModel}
we give an overview of the timing model, in
Section~\ref{sec:likelihood} we write the likelihood function for the
parameters of the stochastic background as well as intrinsic noise
parameters of the pulsars, and introduce the first order approximation
in the amplitude of the cross-correlations, in
Section~\ref{sec:simulations}, we show the effectiveness of our
approximation using simulated gravitational wave backgrounds, and that
the level of bias introduced by our approximation is negligible for
astrophysically reasonable stochastic background amplitudes. We
conclude in Section~\ref{sec:conclusions} with a summary of our
results, compare our results to other work to increase the
computational efficiency of stochastic background
searches~\citep{vh12,Lentati:2012xb,Taylor:2012vx}, and introduce a
technique that can be used to search for a combination of continuous
wave signals and stochastic backgrounds, a possibility suggested by
recent work~\citep{rwh+12}, which will be the basis for future work.

\section{The Timing Model} \label{sec:timingModel}

In pulsar timing we measure the times-of-arrival (TOAs) of radio
pulses emitted from pulsars. These TOAs contain many terms of known
functional form (pulsar period, spin-down, etc.), radiometer noise,
pulse phase jitter, and possibly red noise either from ISM effects,
intrinsic pulsar spin noise~\citep{sc10}, or a stochastic
gravitational wave background (GWB). Let the TOAs for a pulsar be
given by \be t^{\rm obs}=t^{\rm det}(\boldsymbol{\xi}_{\rm true})+n,
\ee where $t^{\rm obs}$ is the observed TOA, $t^{\rm det}$ is the
deterministic modeled TOA parameterized by timing model parameters
$\boldsymbol{\xi}_{\rm true}$, and $n$ is the noise in the measurement
which we will assume to be Gaussian. We will discuss the exact form of
the covariance matrix for the noise $n$ in the next section. Assuming
we have an estimate of the true timing model parameters,
$\boldsymbol{\xi}_{\rm est}$ (either from information gained when
discovering the pulsar or past timing observations), then we can form
the post fit residuals as follows \be \begin{split} \delta t^{\rm
pre}&=t^{\rm obs}-t^{\rm det}(\boldsymbol{\xi}_{\rm est})=t^{\rm
det}(\boldsymbol{\xi}_{\rm true})-t^{\rm det}(\boldsymbol{\xi}_{\rm
est})+n\\ &=t^{\rm det}(\boldsymbol{\xi}_{\rm true})-t^{\rm
det}(\boldsymbol{\xi}_{\rm true}+\delta\boldsymbol{\xi})+n\\
&\approx\left.\frac{\partial t^{\rm det}(\boldsymbol{\xi}_{\rm
true})}{\partial
\delta\boldsymbol{\xi}}\right|_{\delta\boldsymbol{\xi}=0}\delta\boldsymbol{\xi}+n+\mathcal{O}(\delta\boldsymbol{\xi}^{2})\\
&\approx \left.\frac{\partial t^{\rm det}(\boldsymbol{\xi}_{\rm
true})}{\partial
\delta\boldsymbol{\xi}}\right|_{\delta\boldsymbol{\xi}=0}\delta\boldsymbol{\xi}+n\\
&= M\delta\boldsymbol{\xi}+n, \label{eq:pre} \end{split} \ee where $M$
is called the design matrix and we have assumed that our initial
estimate of the model parameters is sufficiently close to the true
values that we can approximate this as a linear system of equations in
$\delta\boldsymbol{\xi}$. In standard pulsar timing analysis, it is
customary to obtain the best fit $\delta\boldsymbol{\xi}$ values
through a weighted least squares minimization of the pre-fit
residuals. In the most general case we should be performing a
\emph{generalized} least squares fit using a general covariance matrix
for the noise $n$; however, in most cases we have no a priori
knowledge of this covariance matrix and therefore assume that it is
just diagonal with elements $\sigma_{i}^{2}$, where $\sigma_{i}$ is
the uncertainty of the $i$th TOA. Previous work \citep{chc+10} has
used an iterative method to estimate the covariance matrix of the
residuals and apply a generalized least squares fit, however; for this
work we will only work with residuals that have been created using a
weighted least squares fit, since that is the standard procedure in
pulsar timing residual generation. The value of chi-squared can be
written in the following way (see \cite{hem06}) \be
\chi^{2}=\sum_{i=1}^{N}\lp \frac{\delta t^{\rm pre}}{\sigma_{i}}
\rp^{2}. \ee Defining $W=1/\sigma_{i}$ we can minimize $\chi^{2}$ \be
\begin{split} 0&=\frac{\partial
\chi^{2}}{\partial\delta\boldsymbol{\xi}}=W^{2}\lp
M\delta\boldsymbol{\xi}+n \rp M^{T}\\ &\Rightarrow
M^{T}W^{2}n=-M^{T}W^{2}M\delta\boldsymbol{\xi}, \end{split} \ee to
obtain our best fit model parameters \be \delta\boldsymbol{\xi}_{\rm
best}=-\lp M^{T}W^{2}M \rp^{-1}M^{T}W^{2}n. \ee Here we have made the
choice to include the weights, $W$, since \textsc{tempo2} does a
weighted fit and we want to reproduce the fitting procedure as
accurately as possible. Finally we obtain the post fit residuals by
substituting the best fit parameters into Eq. \ref{eq:pre} \be
\begin{split}
  r \equiv &\delta t^{\rm post}=M\delta\boldsymbol{\xi}_{\rm best}+n
\\
 &\Rightarrow r=Rn, \end{split} \label{eq:resids} \ee where $r$ is
just shorthand notation for the post-fit residuals and \be
R=\mathbb{I}-M\lp M^{T}W^{2}M \rp^{-1}M^{T}W^{2}, \ee is a an oblique
projection operator that transforms pre-fit to post-fit residuals and
$\mathbb{I}$ is the identity matrix. All of the information about any
noise source or stochastic GWB is encoded in $n$, however; we can
never measure $n$ directly because we must perform the timing model
subtraction. Because of this we seek to work exclusively in terms of
our observable quantities, $r$. It should be noted that  in standard
pulsar timing analysis this process must be iterated. In other words
we form pre-fit residuals from our initial guess of the parameters, we
then minimize the chi-squared to get our best estimates of the
parameters, however this may not be a good fit because we have assumed
that the pre-fit residuals are linear in the parameter offsets. Thus,
we then form new parameter estimates from the best fit parameter
offsets and iterate until the fit converges where the reduced
chi-squared is used as our goodness of fit parameter.

\section{The Likelihood Function} \label{sec:likelihood}

The likelihood function for the timing residuals may be derived very
simply from the likelihood of the underlying pre-fit Gaussian random
processes. In this section we will derive an expression for the
likelihood and introduce our approximation. We will also show that, in
a frequentist sense, the maximum of the expectation value of the
likelihood function is an unbiased estimator of the noise parameters
in the low-signal regime.

Since we have assumed that our noise $n$ is Gaussian and stationary,
for a pulsar timing array with $M$ pulsars we can write the
probability distribution as the multi-variate Gaussian \be
\label{eq:likey} p(\mathbf{n}|\vec\theta)=\frac{1}{\sqrt{\det(2 \pi
\boldsymbol{\Sigma}_{n})}}\exp\lp-\frac{1}{2}\mathbf{n}^T\boldsymbol{\Sigma}_{n}^{-1}\mathbf{n}\rp,
\ee where where \be \mathbf{n}=\bb {n}_{1} \\ {n}_{2}\\ \vdots \\
{n}_{M} \eb \ee is a vector of the noise time-series, $n_{\alpha}(t)$,
for all pulsars, $\boldsymbol{\Sigma}_{n}$ is the \emph{pre}-fit noise
covariance matrix and $\vec\theta$ is a set of parameters that
characterize the noise. However, as we noted above, we do not actually
measure $\mathbf{n}$, we measure the timing residuals
$\mathbf{r}=\mathbf{R}\mathbf{n}$ where \be \label{eq:rmat}
\mathbf{R}=\bb R_{1} & 0 & \hdots & 0\\ 0 & R_{2} & \hdots & 0\\
\vdots & \vdots & \ddots & \vdots\\ 0 & 0 & \hdots & R_{M}\eb. \ee
 We compute the likelihood for $\mathbf{r}$ as follows. Let \be
\label{eq:jacobian}
p(\mathbf{r}|\vec\theta)d\mathbf{r}=p(\mathbf{n}|\vec\theta)d\mathbf{n}\Rightarrow
p(\mathbf{r}|\vec\theta)=p(\mathbf{n}|\vec\theta)\left|\frac{d\mathbf{n}}{d\mathbf{r}}\right|,
\ee where $|\cdot|$ represents the determinant. We evaluate the
Jacobian by \emph{assuming} that $\mathbf{R}$ is invertible and
writing $\mathbf{n}=\mathbf{R}^{-1}\mathbf{r}$, therefore \be
\left|\frac{d\mathbf{n}}{d\mathbf{r}}\right|=\left|\mathbf{R}^{-1}\right|=\frac{1}{\left|\mathbf{R}\right|}=\frac{1}{\sqrt{\det(\mathbf{R}\mathbf{R}^T)}}.
\ee Substituting this result into Eq. \ref{eq:jacobian} we obtain \be
p(\mathbf{r}|\vec\theta)=\frac{1}{\sqrt{\det(2\pi
\mathbf{R}\boldsymbol{\Sigma}_{n}\mathbf{R}^T)}}\exp\lp-\frac{1}{2}\mathbf{r}(\mathbf{R}^{-1})^T\boldsymbol{\Sigma}_{n}^{-1}
\mathbf{R}^{-1}\mathbf{r}\rp. \ee The product
$\mathbf{R}\boldsymbol{\Sigma}_{n}\mathbf{R}^T$ is just the covariance
matrix for the residuals \be \boldsymbol{\Sigma} = \langle \mathbf{r}
\mathbf{r}^T \rangle = \mathbf{R}\langle \mathbf{n} \mathbf{n}^T
\rangle \mathbf{R}^T= \mathbf{R}\boldsymbol{\Sigma}_{n}\mathbf{R}^{T},
\ee so that the likelihood in terms of the timing residual data is
simply \be \label{eq:liker}
p(\mathbf{r}|\vec\theta)=\frac{1}{\sqrt{\det(2 \pi
\boldsymbol{\Sigma})}}\exp\lp-\frac{1}{2}\mathbf{r}^T\boldsymbol{\Sigma}^{-1}\mathbf{r}\rp.
\ee The inverse of $\boldsymbol{\Sigma}$ does not formally exist since
we have removed degrees of freedom by fitting out the timing model. In
practice, we can make use of a singular value decomposition to compute
the determinant and pseudoinverse to evaluate the likelihood. Viewed
in this way, the likelihood function for the residuals is simply a
change of coordinates where $\mathbf{R}$ is a linear (but not
invertible) map from $\mathbf{n}\rightarrow
\mathbf{r}=\mathbf{R}\mathbf{n}$.

The covariance matrix for the timing residuals is the block matrix,
\be \label{eq:cov} \boldsymbol{\Sigma}=\bb P_{1} & S_{12} & \hdots &
S_{1M}\\ S_{21} & P_{2} & \hdots & S_{2M}\\ \vdots & \vdots & \ddots &
\vdots\\ S_{M1} & S_{M2} & \hdots & P_{M}\eb, \ee where \begin{align}
P_{{\alpha}}&=\langle r_{\alpha}r_{\alpha}^{T}\rangle,\\
S_{\alpha\beta}&=\langle
r_{\alpha}r_{\beta}^{T}\rangle\big|_{\alpha\ne \beta}, \end{align} are
the auto-covariance and cross-covariance matrices, respectively, for
each set of residuals. It is very important to note that we work
\emph{exclusively} in the post-fit variables. As above we use the
post-fit residuals, $r_\alpha=R_\alpha n_\alpha$ and the post-fit
auto- and cross-correlation matrices,
$P_{\alpha}=R_{\alpha}P_{\alpha}^{\rm prefit}R_{\alpha}^{T}$ and
$S_{\alpha\beta}=R_{\alpha}S_{\alpha\beta}^{\rm prefit}R_{\beta}^{T}$.
Henceforth, we will drop any mention of pre-fit or post-fit as we will
only work with post-fit variables.

It is worth pointing out that this treatment is somewhat different
from previous Bayesian analyses \citep{hlm+09,vl10,vhj+11} (VHML). We
use a \emph{conditional} pdf whereas VHML used a \emph{marginalized}
pdf. In other words, we fix the best fit parameter offsets,
$\delta\boldsymbol{\xi}_{\rm best}$ through our use of the projection
matrix $R$, whereas VHML marginalizes over the parameter offsets
$\delta\boldsymbol{\xi}$ (See Appendix \ref{app:likelihood} for more
details).

We would like to use the likelihood to determine the spectral index,
$\gamma_{\rm gw}$, and amplitude, $A_{\rm gw}$, of the stochastic
background from our data. The GW parameters are the same for all
pulsars. In addition, each pulsar will have intrinsic noise parameters
as well. The intrinsic pulsar timing noise is normally parametrized
with four parameters: an amplitude $A_{\alpha}$ and spectral index
$\gamma_{\alpha}$ for a power law red noise process, and EFAC and
EQUAD parameters, $\mathcal{F}_{\alpha}$ and $\mathcal{Q}_{\alpha}$,
for white noise processes. In general the EFAC parameter is a
multiplicative factor representing any systematic effects in the
uncertainty in each TOA based on the cross correlation of the folded
pulse profile with a template \citep{twdw92}. The EQUAD parameter is
an extra white noise parameter that is added to the TOA error in
quadrature and could represent the expected pulse phase jitter
\citep{cs10} and other white noise processes that are un-accounted
for. Therefore, we write our auto-covariance as a sum of a common GWB
term and a pulsar dependent term \be
P_{\alpha}=N_{\alpha}+S_{a\alpha}, \ee where $N_{\alpha}$ is the
intrinsic noise auto-covariance matrix and $S_{a\alpha}$ is the common
GWB auto-covariance matrix for pulsar $\alpha$. It is convenient to
work in a block matrix notation where \be
\boldsymbol{\Sigma}=\mathbf{N}+\mathbf{S}_{a}+\mathbf{S}_{c}=\mathbf{P}+\mathbf{S}_{c},
\ee where $\mathbf{P}$ is a block diagonal matrix with diagonals
$P_{\alpha}$ and $\mathbf{S}_{c}$ is block matrix with off diagonals
$S_{\alpha\beta}$, and zero block matrices on the diagonal.

We will now quickly show that, in a frequentist sense, the maximum of
the expectation value of the likelihood function is an unbiased
estimator of our signal parameters $\vec\theta=\{A_{\rm
gw},\gamma_{\rm
gw},A_{\alpha},\gamma_{\alpha},\mathcal{F}_{\alpha},\mathcal{Q}_{\alpha}\}$.
We write the log likelihood function as \be
\ln\,\mathcal{L}=-\frac{1}{2}\left[ \Tr\, \ln \boldsymbol{\Sigma}
+\mathbf{r}^{T}\boldsymbol{\Sigma}^{-1}\mathbf{r} \right], \ee where
we have used the fact that $\ln \det(A)=\Tr \ln(A)$ for a general
matrix, $A$. To show that the maximum of the expectation value of this
likelihood function is an unbiased estimator of the signal parameters,
$\vec\theta$, we wish to show that it is maximized, on average, for
signal parameters $\vec\theta=\vec\theta_{\rm true}$. Taking the
expectation value we obtain \be
\langle\ln\,\mathcal{L}\rangle=-\frac{1}{2}\Tr \left[ \ln
\boldsymbol{\Sigma} +\mathbf{X}\boldsymbol{\Sigma}^{-1} \right], \ee
where $\mathbf{X}=\langle \mathbf{r}\mathbf{r}^{T} \rangle$ is the
covariance matrix of the data. Defining
$\partial_{i}=\partial/\partial\theta_{i}$ we obtain \be
\partial_{i}\langle\ln\,\mathcal{L}\rangle=-\frac{1}{2}\Tr \left[
\boldsymbol{\Sigma}^{-1}\partial_{i} \boldsymbol{\Sigma}
-\mathbf{X}\boldsymbol{\Sigma}^{-1}\partial_{i}\boldsymbol{\Sigma}\boldsymbol{\Sigma}^{-1}
\right]. \ee Assuming that our noise model is correct, we have
$\mathbf{X}=\boldsymbol{\Sigma}$ and \be
\partial_{i}\langle\ln\,\mathcal{L}\rangle=-\frac{1}{2}\Tr \left[
\boldsymbol{\Sigma}^{-1}\partial_{i} \boldsymbol{\Sigma}
-\partial_{i}\boldsymbol{\Sigma}\boldsymbol{\Sigma}^{-1} \right]=0,
\ee where we have used the fact that $\Tr(AB)=Tr(BA)$ for general
matrices, $A$ and $B$. Therefore, the maximum of the expectation value
of the likelihood function is an unbiased estimator of our model
parameters $\vec\theta$.

\subsection{Likelihood with first order approximation}

In practice the matrix $\boldsymbol{\Sigma}$ is quite large and
therefore, computationally prohibitive to invert. Since many
multi-frequency residual datasets now have on the order of $10^{3}$
points, for many modern PTAs the matrix $\boldsymbol{\Sigma}$ will be
of order $10^{4}\times 10^{4}$. We would like to avoid inverting the
full covariance matrix if at all possible. First let us rewrite the
cross-covariance as
$\mathbf{S}_{c,\alpha\beta}=\zeta_{\alpha\beta}\mathbf{S}_{\alpha\beta}$,
where $\mathbf{S}_{\alpha\beta}$ is the temporal cross covariance
between pulsar $\alpha$ and pulsar $\beta$. The coefficients represent
the spatial correlations and are given by the Hellings and Downs
coefficients \be \begin{split}
\zeta_{\alpha\beta}&=\frac{3}{2}\frac{1-\cos\xi_{\alpha\beta}}{2}\ln\left(\frac{1-\cos\xi_{\alpha\beta}}{2}\right)\\
&-\frac{1}{4}\frac{1-\cos\xi_{\alpha\beta}}{2}+\frac{1}{2}+\frac{1}{2}\delta_{\alpha\beta},
\end{split} \ee where $\xi_{\alpha\beta}$ is the angular separation of
pulsars $\alpha$ and $\beta$, and $\delta_{\alpha\beta}$ is the
Kronecker delta. We denote
$\mathbf{P}=\delta_{\alpha\beta}\mathbf{P}_{\alpha\beta}$ as the
auto-covariance matrix of pulsar $\alpha$ describing the noise and
auto-covariance of the GWB. We then use the following notation to form
matrices from indexed quantities: $\mathbf{P}=\{P_{\alpha\beta}\}$.
Now, we perform the expansion of $\boldsymbol{\Sigma}^{-1}$ in terms
of the coefficients $\zeta_{\alpha\beta}$ \be \begin{split}
\boldsymbol{\Sigma}^{-1}&=\lp
\mathbf{P}+\{\zeta_{\alpha\beta}\mathbf{S}_{\alpha\beta}\}\rp^{-1}=\lp
\mathbb{I}+\mathbf{P}^{-1}\{\zeta_{\alpha\beta}\mathbf{S}_{\alpha\beta}\}
\rp^{-1}\mathbf{P}^{-1}\\ &\approx \mathbf{P}-\left\{
\sum_{\beta,\mu}\zeta_{\beta\mu}\mathbf{P}_{\alpha\beta}^{-1}\mathbf{S}_{\beta\mu}\mathbf{P}_{\mu\nu}^{-1}\right\}\\
&+\left\{ \sum_{\beta,\mu,\nu}\zeta_{\beta\mu}\zeta_{\mu\nu}
\mathbf{P}_{\alpha\beta}^{-1}\mathbf{S}_{\beta\mu}\mathbf{P}_{\mu\mu}^{-1}\mathbf{S}_{\mu\nu}\mathbf{P}_{\nu\sigma}^{-1}\right\}+\mathcal{O}(\zeta^{3}).
\end{split} \ee It is also possible to expand the determinant term in
a similar fashion \be \begin{split} \ln
\det\boldsymbol{\Sigma}&=\Tr\ln\boldsymbol{\Sigma}=\Tr\ln(\mathbf{P}+\{\zeta_{\alpha\beta}\mathbf{S_{\alpha\beta}}\})\\
&=\Tr\left[\ln\mathbf{P}+\ln(\mathbb{I}+\mathbf{P}^{-1}\{\zeta_{\alpha\beta}\mathbf{S}_{\alpha\beta}\})\right]\\
&\approx
\Tr\Bigg[\ln\mathbf{P}+\mathbf{P}^{-1}\{\zeta_{\alpha\beta}\mathbf{S}_{\alpha\beta}\}\\
&-\left\{ \sum_{\beta,\mu,\nu}\zeta_{\beta\mu}\zeta_{\mu\nu}
\mathbf{P}_{\alpha\beta}^{-1}\mathbf{S}_{\beta\mu}\mathbf{P}_{\mu\mu}^{-1}\mathbf{S}_{\mu\nu}\mathbf{P}_{\nu\sigma}^{-1}\right\}\Bigg]+\mathcal{O}(\zeta^{3}).
\end{split} \ee Here, the order $\mathcal{O}(\zeta)$ term is zero
because $\mathbf{P}$ is block diagonal and
$\{\mathbf{S}_{\alpha\beta}\}$ is block traceless and the trace of the
product of a diagonal matrix and traceless matrix vanishes. If we
ignore all terms of $\zeta^{2}$ and higher order and return to our
original notation then we see that \begin{align} \label{eq:invExp}
&\boldsymbol{\Sigma}^{-1}\approx
\mathbf{P}^{-1}-\mathbf{P}^{-1}\mathbf{S}_{c}\mathbf{P}^{-1}+\mathcal{O}(\zeta^{2})\\
\label{eq:detExp}
&\ln\det\boldsymbol{\Sigma}\approx\Tr\ln\mathbf{P}+\mathcal{O}(\zeta^{2}).
\end{align} This derivation may give us the sense that this expansion
may hold true for all GWB amplitudes; however, this is not true as we
will now show. Although we have written this approximation in terms of
an expansion in the Hellings and Downs coefficients, it is also useful
to think of it as an expansion in the amplitude of the GWB. Indeed,
that it how it was conceived of in~\cite{abc+09}. We have not
performed a true first order expansion however, since the inverse of
the auto-correlations matrix $\mathbf{P}^{-1}=(\mathbf{N}+A_{\rm
gw}^{2}\mathbf{A}_{a})^{-1}$ contains terms of infinite order in the
amplitude. We can essentially think of the $\mathcal{O}(\zeta)$ terms
in Equations \ref{eq:invExp} and \ref{eq:detExp} as the corrections to
the amplitude parameter when we have a spatially correlated signal.
Thus, we have truncated these correction terms at $\mathcal{O}(A_{\rm
gw}^{2})$ and we would not expect this approximation to hold as
$A_{\rm gw}$ becomes large with respect to the intrinsic noise in the
pulsar as we will show in Section \ref{sec:simulations}. With these
approximations, it is now possible to write the approximate
log-likelihood \be \begin{split} \ln \mathcal{L}&=-\frac{1}{2}\left[
\Tr\ln \mathbf{P}
+\mathbf{r}^{T}\mathbf{P}^{-1}\mathbf{r}-\mathbf{r}^{T}\mathbf{P}^{-1}\mathbf{S}_{c}\mathbf{P}^{-1}\mathbf{r}
\right]\\ &=-\frac{1}{2}\sum_{\alpha=1}^M \bigg[\Tr\ln P_{\alpha}
+r_{\alpha}^TP_{\alpha}^{-1}r_{\alpha}\\
 &-\sum_{\beta\ne\alpha}^Mr_{\alpha}^T
P_{\alpha}^{-1}S_{\alpha\beta}P_{\beta}^{-1}r_{\beta}\bigg].
 \label{eq:foLike} \end{split} \ee In the second line we have
explicitly written out the sum over pulsars and pulsar pairs in order
to highlight the fact that we only need to invert the individual
auto-covariance matrices as opposed to the inverse of the full block
covariance matrix, thereby, significantly reducing the computational
cost of a single likelihood evaluation. Consider a PTA with $M$
pulsars with $N$ TOAs each. For a full likelihood evaluation we must
perform one Cholesky inversion of the full covariance matrix which
scales like $\sim \alpha(MN)^{3}$ and $\sim M^{2}$ matrix
multiplications which scale like $\sim\beta N^{3}$. However, one
evaluation of the first order likelihood requires $M$ Cholesky
inversions which scale like $\sim \alpha N^{3}$ and $M$ matrix
multiplications which, again, scale like $\sim \beta N^{3}$. Though
benchmarking tests we have found that $\beta\sim 10\alpha$ and thus
the matrix multiplications will dominate both likelihood calls for a
reasonable sized PTAs ($M\lesssim100$) resulting in a computation
speedup factor of $\sim (\alpha/\beta) M^{2}$.


It is possible to analytically show that the maximum of the
expectation value of this approximate likelihood is an unbiased
estimator in the same manner as above. First we take the expectation
value of the log-likelihood \be
\langle\ln\mathcal{L}\rangle=-\frac{1}{2}\Tr\left[ \ln \mathbf{P}
+\mathbf{X}\mathbf{P}^{-1}-\mathbf{X}\mathbf{P}^{-1}\mathbf{S}_{c}\mathbf{P}^{-1}\right]
\ee and then take a derivative with respect to a model parameter \be
\begin{split}
\partial_{i}\langle\ln\mathcal{L}\rangle&=-\frac{1}{2}\Tr\bigg[\mathbf{P}^{-1}\partial_{i}\mathbf{P}-\mathbf{X}\mathbf{P}^{-1}\partial_{i}\mathbf{P}\mathbf{P}^{-1}\\
&+\mathbf{X}\mathbf{P}^{-1}\partial_{i}\mathbf{P}\mathbf{P}^{-1}\mathbf{S}_{c}\mathbf{P}^{-1}
-\mathbf{X}\mathbf{P}^{-1}\partial_{i}\mathbf{S}_{c}\mathbf{P}^{-1}\\
 &+\mathbf{X}\mathbf{P}^{-1}\mathbf{S}_{c}\mathbf{P}^{-1}\partial_{i}\mathbf{P}\mathbf{P}^{-1}\bigg].
\end{split} \ee Here we will work in the small signal regime where
$A_{\rm gw}^{2}$ is small compared to the amplitude of the intrinsic
noise. Assuming that we have modeled the covariance matrix correctly,
we have $\mathbf{X}=\boldsymbol{\Sigma}$. Writing out the explicit
amplitude dependence we assume \begin{align}
\mathbf{P}&=\mathbf{N}+A_{\rm gw}^{2}\mathbf{A}\Rightarrow
\mathbf{P}^{-1}\approx\mathbf{N}^{-1}-A_{\rm
gw}^{2}\mathbf{N}^{-1}\mathbf{A}\mathbf{N}^{-1}\\
\boldsymbol{\Sigma}&=\mathbf{N}+A_{\rm gw}^{2}\mathbf{A}+A_{\rm
gw}^{2}\mathbf{C}, \end{align} where $\mathbf{N}$, $\mathbf{A}$, and
$\mathbf{C}$ are the auto-covariance of the noise, the auto-covariance
of the GWB and the cross-covariance of the GWB, respectively. Then, to
first order in $A_{\rm gw}^{2}$ we have \be \begin{split}
\partial_{i}\langle\ln\mathcal{L}\rangle&=-\frac{1}{2}\Tr\bigg[
\mathbf{N}^{-1}\partial_{i}(A_{\rm gw}^{2} \mathbf{A})\\ &-
\mathbf{N}^{-1}\partial_{i}(A_{\rm gw}^{2} \mathbf{A})
-\partial_{i}(A_{\rm gw}^{2} \mathbf{C})\mathbf{N}^{-1}\bigg]=0,
\end{split} \ee where the first two terms cancel and the third term is
the trace of the product of a diagonal matrix and a traceless matrix.
Thus, to first order in $A_{\rm gw}^{2}$, the maximum of the
expectation value of this approximate likelihood is an unbiased
estimator of the our signal parameters $\theta$ in the weak signal
limit.

\section{Simulations} \label{sec:simulations}

Here we will compare our first order likelihood approximation to the
full likelihood of VHML and perform mock searches of simulated data
with and without an injected stochastic GWB in order to demonstrate
its efficacy. We will also perform monte-carlo simulations to test the
consistency of our likelihood function. These simulations are solely
meant as a proof of principle and do not claim to reproduce all
features of real PTA data (irregular sampling, jumps, time varying DM
corrections, etc.). However, our analysis method makes no assumptions
about sampling by operating in the time domain and takes all timing
model parameters into account via the projection matrices introduced
in Section \ref{sec:timingModel}. The application of this method to
real NANOGrav and IPTA datasets will be the subject of future work.
For all simulations in the present work we use \textsc{tempo2} and and
the \texttt{fake}, \texttt{GWbkgrd}, \texttt{general2} and
\texttt{designmatrix} plugins to generate the residuals and the
corresponding design matrices. All simulated white noise is solely
radiometer noise at the level of 100 ns unless otherwise noted.

\subsection{Mock searches} \label{sec:mockSearch}

\begin{figure*}[t]
  \begin{center}
  \subfigure[]{
  	\includegraphics[scale=0.95]{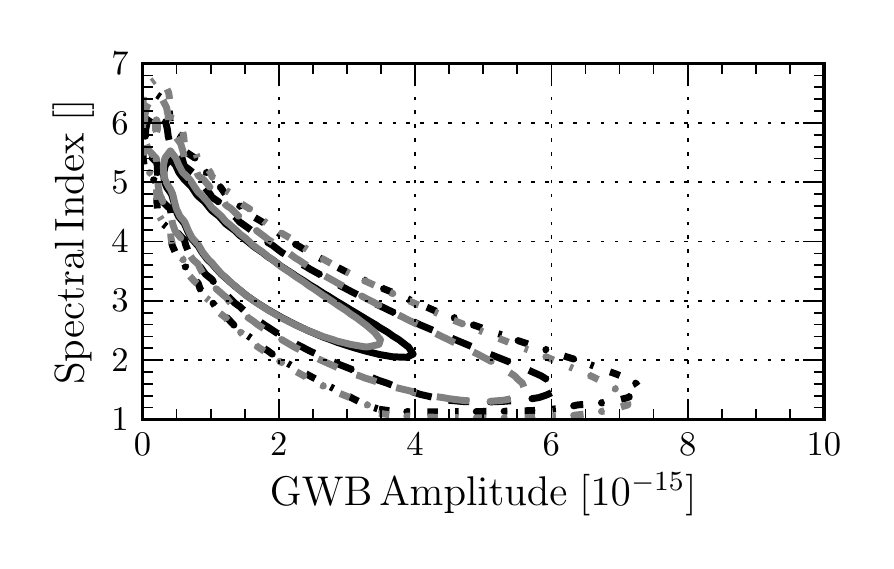}
\label{fig:compareA}}
  \subfigure[]{
  	\includegraphics[scale=0.95]{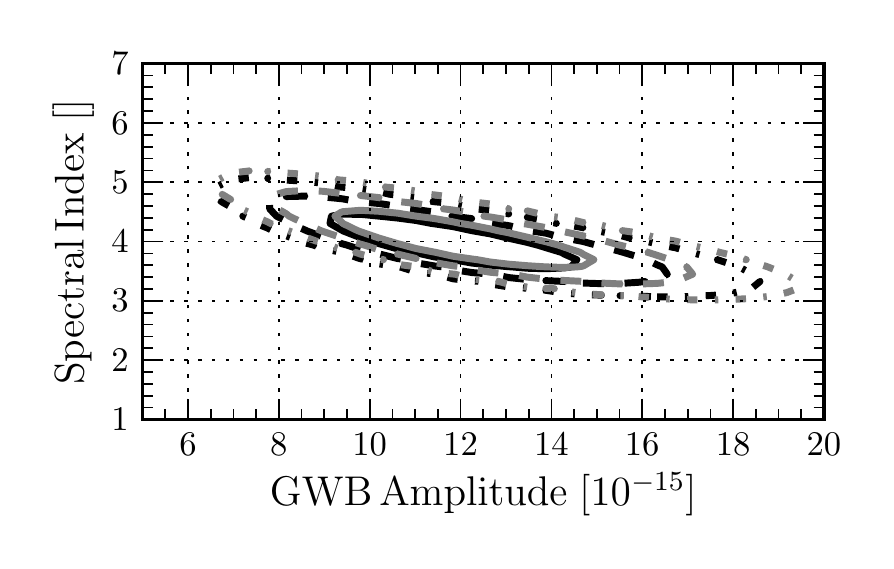}
\label{fig:compareB}}\\
   \end{center}
  \caption{Comparison of full likelihood (gray) of \citet{hlm+09} and
the first order likelihood (black). (a): 10 pulsars $A=1\times
10^{-15}$, (b): 10 pulsars $A=1\times 10^{-14}$ } \label{fig:compare}
\end{figure*}

First we will perform a simple test to compare the first order
likelihood of this work and the full likelihood of VHML. Here we use a
PTA with 10 pulsars observed at a cadence of 20 TOAs per year for 5 years
 where we have fixed the EFAC parameter to be one
(all white noise is encompassed in error bars as simulated) and assume
that there is no intrinsic red noise, resulting in a search over two
parameter; the amplitude of the stochastic GWB, $A$, and the power
spectral index, $\gamma$. For both cases a grid search was carried out
with 100 points in each dimension and $A\in (0,1\times 10^{-14})$ for
an injected value of $A=1\times 10^{-15}$ and $A\in (0,2\times
10^{-14})$ for an injected value of $A=1\times 10^{-14}$, all the
while we have $\gamma \in [1,7]$. The results are presented in Figure
\ref{fig:compare} where the contours denote the one, two and three
sigma credible regions, the gray contours are from the VHML likelihood
function and the black contours are from the first order likelihood.
 In Figure \ref{fig:compareA} we have injected a stochastic GWB with
$A=1\times 10^{-15}$ and $\gamma=13/3$. First we notice that the
injected value ('$\times$' marker) is well within the 1-sigma credible
regions for both likelihood functions. We also see that the confidence
contours are nearly identical, with the first order likelihood
preferring slightly larger amplitudes and smaller spectral indices.
This simulation indicates that the first order likelihood is a very
good approximation to the full likelihood when our signal is
relatively small, showing no discernible bias and faithfully
reproducing nearly identical credible regions.
 
 In Figure \ref{fig:compareB} we have injected a stochastic GWB with
$A=1\times 10^{-14}$ and $\gamma=13/3$. Again, the injected value lies
within the 1-sigma credible region, however; now we do notice a
difference between two credible regions from the full and first order
likelihoods. The first order likelihood is biased towards lower
amplitudes and lower spectral indices. In fact we can almost see where
the first order approximation begins to break down. Notice that the
contours are nearly identical for lower amplitudes and deviate more
with increasing amplitude. This behavior is not surprising in that we
know that this likelihood is only unbiased to first order in the
amplitude as shown in Section \ref{sec:likelihood}. In fact, it is
impressive that this approximation performs this well with only a
small bias in the large signal limit (even with timing residuals lower
than 100 ns in many pulsars, the signal-to-noise-level of the data
simulated here is well above any reasonable estimates for future PTA
sensitivities.). This bias will be discussed further in Section
\ref{sec:edf}.
 
 The simulations used in the work have been quite ideal and do not
contain any systematic effects such as clock errors which can manifest
as a correlated noise source with uniform correlation coefficients
\citep{ych+11}, errors in solar system ephemerides, which can manifest as dipole signals in the residuals, or
new physics such as non-gr polarization modes \citep{ljp08,ss12} or
massive gravitons \citep{ljp+10} which would change the shape of the
Hellings and Downs curve. We have, for the most part, also assumed
that the intrinsic pulsar noise can be assumed to be white gaussian
noise with no discernible red noise. While previous work suggests that
there will be red noise present in many MSPs \citep{sc10}, analyses of
the present timing data \citep{vhj+11,delphine,esd+13} suggest that
the data is white noise dominated and there is little to no evidence
for red noise. However further study of the model selection problem
taking in to account the aforementioned effects is crucial to present
detection efforts and will be the subject of a future paper.

\subsection{The detection problem} 

We now turn to the question of
detection. In a Bayesian analysis we would like to compute the odds
that there is a GWB present in our data. Not surprisingly, the tool
normally used to this end is the Odds ratio of Bayes factors. Consider
two models that we will label $M_{1}$ and $M_{2}$, then the Odds ratio
is defined as \be
\mathcal{O}=\mathcal{B}(M_{1},M_{2}|\mathbf{r})\frac{p(M_{1})}{p(M_{2})},
\ee
 where
 \be
 \mathcal{B}(M_{1},M_{2}|\mathbf{r})=\frac{\int
d\vec\theta_{1}\,p(\mathbf{r}|\vec\theta_{1},M_{1})p(\vec\theta_{1})}{\int
d\vec\theta_{2}\,p(\mathbf{r}|\vec\theta_{2},M_{2})p(\vec\theta_{2})}
 \ee
 is the Bayes factor (i.e the ratio of the marginalized likelihood
functions over parameters $\vec\theta_{1}$ and $\vec\theta_{2}$
corresponding to models $M_{1}$ and $M_{2}$ respectively),
$\mathbf{r}$ is our data and $p(M_{1})$ and $p(M_{2})$ are the a
priori probabilities on models $M_{1}$ and $M_{2}$ respectively. Note
that the Bayes factor is the data dependent part of the odds ratio
where the a priori probabilities of the models is somewhat subjective,
and as such, we will only consider Bayes factors when discussing
detection in the this work \footnote{It is possible to use
astrophysical information such as the expected level of the stochastic
background compared to our noise or the expectation number of single
sources to construct the a priori probabilities. Here we will quantify
our ignorance by considering equal a priori probabilities of all
tested models.}. For our purposes, we would like to compare at least
three different models when weighing the odds of a stochastic GWB in
our data:
 \begin{enumerate}
 
 \item $M_{\rm gw}$: A power law stochastic GWB with spatial
correlations described by the Hellings and Downs coefficients
$\zeta_{\alpha\beta}$, amplitude $A_{\rm gw}$ and power spectral index
$\gamma_{\rm gw}$, individual power law red noise processes for each
pulsar with amplitude $A_{\alpha}$ and power spectral index
$\gamma_{\alpha}$ and white noise for each pulsar characterized by an
EFAC parameter $\mathcal{F}_{\alpha}$ and EQUAD parameter
$\mathcal{Q}_{\alpha}$.
 
 \item $M_{\rm corr}$: A common red noise process among pulsars (as
suggested in \cite{sc10}) with no spatial correlations and individual
intrinsic red and white components as in model $M_{\rm gw}$.
 
 \item $M_{\rm null}$: Only intrinsic red and white noise processes
with no common red or white noise components among pulsars.
 
 \end{enumerate}
 Comparing models $M_{\rm gw}$ and $M_{\rm null}$ will tell us whether
or not there is evidence for any common red noise in our data but it
will not necessarily tell us that this common noise is due to the
stochastic GWB or some other common red noise source. Hence, a large
Bayes factor $\mathcal{B}(M_{\rm gw},M_{\rm null}|\mathbf{r})$ is
necessary but not sufficient for detection. However, the comparison of
models $M_{\rm gw}$ and $M_{\rm corr}$ can really give us information
about the nature of the common red noise signal. As the two
aforementioned models are identical except for the spatial
correlations, a large Bayes factor $\mathcal{B}(M_{\rm gw},M_{\rm
corr}|\mathbf{r})$ will give us the odds that there is a common red
noise process described spatial correlations $\zeta_{\alpha\beta}$.
Since these \emph{spatial} correlations are the signature of a
stochastic GWB, the condition that this Bayes factor be large is both
the necessary and sufficient condition for detection. In fact, this
Bayes factor is closely related to signal-to-noise ratios in previous
detection schemes \citep{jhl+05,abc+09,ych+11,cce+12} that measure the
significance of the cross correlations.
 
This first order likelihood approximation has already been tested on
the open and closed \citep{esc12b} IPTA Mock Data Challenge, where all
challenges consisted of 130 data points per pulsar with 36 pulsars.
For the closed data challenge, we have computed the Bayes factors
mentioned in the previous section. In \cite{esc12b} we have shown that
we do indeed see very strong evidence for both a common red noise
signal and a red noise signal with spatial correlations described by
the Hellings and Downs coefficients. However, as we mentioned above,
although in this case, the evidence for both models $M_{\rm gw}$ and
$M_{\rm corr}$ is very high, as we expect, the Bayes factor
$\mathcal{B}(M_{\rm gw},M_{\rm null})$ is much larger than
$\mathcal{B}(M_{\rm gw},M_{\rm corr})$. For this reason, we expect
that in analysis of real PTA data we will begin to see strong evidence
for common red noise before we are able to see strong evidence for the
expected cross correlations. In other words, as we gain more
sensitivity, the first two terms in Eq. \ref{eq:foLike} will dominate
the likelihood function and the third term will only play a
significant role as our sensitivity increases further. A full analysis
of this feature along with projected sensitivity curves based on
future pulsar timing campaigns and hardware upgrades will be explored
in future work.

closed MDCs. The open datasets, in particular open MDC1, act as
illustrative cases where the first-order approximation breaks down and
shows a large bias in parameter estimation. (Show plot or Rutger's
results along with ours).

\subsection{The Empirical Distribution Function} \label{sec:edf}

Here we will test the consistency and unbiasedness of our model
through injections. Simply put, it is a type of hypothesis testing
similar to the Kolmogorov-Smirnov test. In this test the
null-hypothesis, our analysis method is internally consistent, is
accepted when for $x$\% of realizations, the true injected parameter
lies within the inner $x$\% of the marginalized posterior
distribution. A similar test was done recently in \cite{vhl12} in
\emph{one} dimension through the use of the empirical distribution
function (EDF). Here we will review this method and generalize it to
two dimensional marginalized posterior distributions. We define the
inner high-probability region (HPR) of the two-dimensional
marginalized posterior distribution as \be \begin{split}
\int_{W}p(\theta_{1},\theta_{2})d\theta_{1}d\theta_{2}&=a\\ W&=\{
\theta_{1},\theta_{2}\in \mathbb{R} :
p(\theta_{1},\theta_{2})>L_{a}\}, \label{eq:hpr} \end{split} \ee where
$L_{a}$ is some value $>0$ unique to each $a$ that corresponds to a
curve of equal probability in the two-dimensional parameter space. In
practice we lay down a grid in this two-dimensional parameter space
and perform our search over the two parameters of interest (for the
stochastic background we search over $A$ and $\gamma$, the
dimensionless strain amplitude and power spectral index of the GWB).
We then define a set of points $\{A_{i},\gamma_{i}\}\in
\mathcal{S}_{a} : p(A_{i},\gamma_{i})>L_{a}$, that is to say we find
all points in our grid that correspond to posterior values that lie
inside our contour curve $L_{a}$. To determine if the injected values
of $\{A_{\rm true},\gamma_{\rm true}\}$ lie within the HPR we simply
check to see if the injected values are consistent with the set
$\mathcal{S}_{a}$. To do this we first define the complementary set to
be $\bar{\mathcal{S}}_{a}$ such that points that are in this set are
outside or the HPR. Now we define two chi-squared functions in the
parameter space \begin{align}
\chi_{a}(A_{i},\gamma_{i})^{2}&=\left(\frac{A_{i}-A_{\rm true}}{A_{\rm
true}}\right)^{2}+\left(\frac{\gamma_{i}-\gamma_{\rm
true}}{\gamma_{\rm true}}\right)^{2}\\
\bar{\chi}_{a}(A_{j},\gamma_{j})^{2}&=\left(\frac{A_{j}-A_{\rm
true}}{A_{\rm true}}\right)^{2}+\left(\frac{\gamma_{j}-\gamma_{\rm
true}}{\gamma_{\rm true}}\right)^{2}, \end{align} where
$\{A_{i},\gamma_{i}\}$ and $\{A_{j},\gamma_{j}\}$ are elements of the
sets $\mathcal{S}_{a}$ and $\bar{\mathcal{S}}_{a}$, respectively.
Finally, we define the empirical distribution function (EDF) as \be
F_{k}(a)=\frac{1}{k}\sum_{n=1}^{k}\Theta(\min \bar{\chi}_{a}^{2}-\min
\chi_{a}^{2}), \ee where $\Theta(x)$ is the Heaviside function. The
term inside the sum indicates an event when the injected values are
``closer'' (in the chi-squared sense) to one of the elements of
$\mathcal{S}_{a}$ than to any of the elements of
$\bar{\mathcal{S}}_{a}$, therefore we can say that the values
$\{A_{\rm true},\gamma_{\rm true}\}$ join the set $\mathcal{S}_{a}$
and lie within the HPR defined in Eq. \ref{eq:hpr}. Now that we have
defined our EDF, the rest of the analysis mimics \cite{vhl12}.

\begin{figure}[!h]
  \begin{center}
\includegraphics[width=\columnwidth]{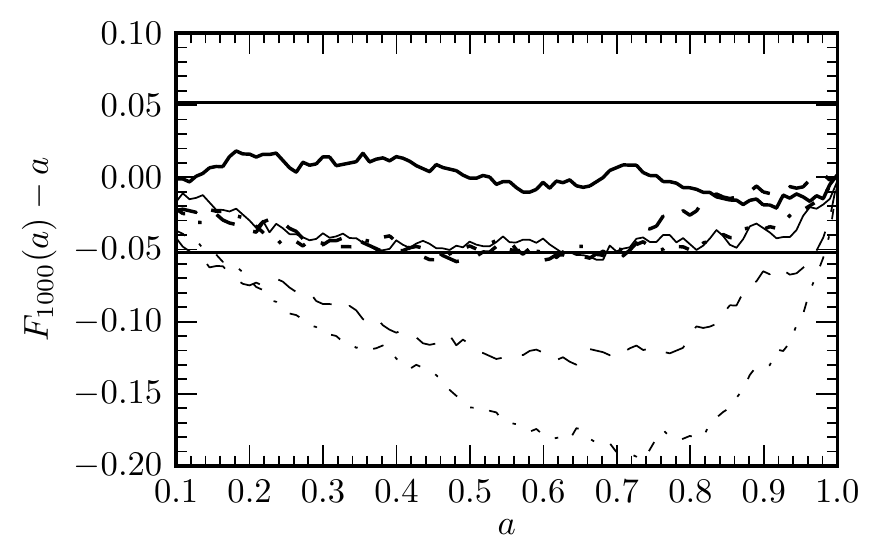}
  \end{center}
  \caption{Empirical distribution function for 6 scenarios. The thick
lines denote a 10 pulsar PTA and the thin lines denote a 15 pulsar PTA
and the solid, dashed and dotted lines denote injected stochastic GWB
amplitudes of $1\times 10^{-15}$, $2\times 10^{-15}$, and $3\times
10^{-15}$, respectively. The solid lines at $\pm 0.052$ represent the
value at which we should reject the null-hypothesis that our analysis
method is consistent and unbiased. } \label{fig:combined_edf}
\end{figure}

For this analysis we simulated 1000 datasets for 6 different
scenarios. In all cases we chose the white noise level to be 100 ns
while we chose GWB amplitudes of $1\times 10^{-15}$, $2\times
10^{-15}$, and $3\times 10^{-15}$ for PTAs with both 10 and 15 pulsars
with a 5 year baseline. Figure \ref{fig:combined_edf} shows the EDF
for the six models outlined above.
The thick lines denote a 10 pulsar PTA and the thin lines denote a 15
pulsar PTA and the solid, dashed and dotted lines denote injected
stochastic GWB amplitudes of $1\times 10^{-15}$, $2\times 10^{-15}$,
and $3\times 10^{-15}$, respectively. The solid lines at $\pm 0.052$
represent the value at which we should reject the null-hypothesis that
our analysis method is consistent and unbiased. Firstly, we note that
for both the 10 and 15 pulsar PTA, our analysis method is consistent
for an injected amplitude of $A=1\times 10^{15}$. We obtain similar results in the 10 pulsar case for amplitudes
of $A=2\times 10^{-15}$ and $A=3\times 10^{-15}$. Here we do see that
our method is indeed slightly biased for these larger amplitudes but
the degree of bias is almost negligible. However, for these same
amplitudes in the 15 pulsar case there is a significant bias. Even
though there is a bias present in these scenarios, the EDF does not
give information about how this bias presents itself in the two
dimensional parameter space.
\begin{figure*}[!t]
  \begin{center}
  \includegraphics[scale=1.0]{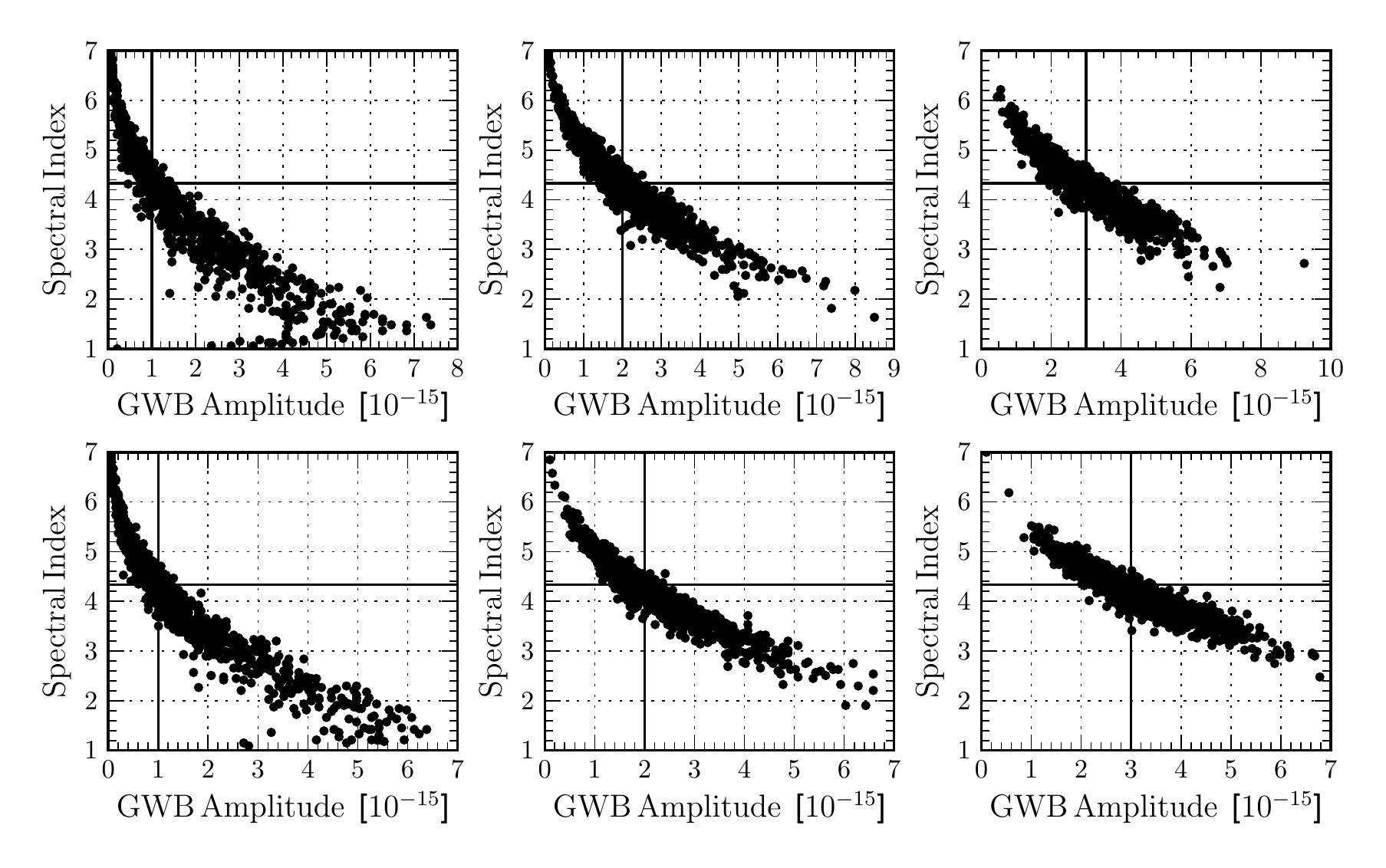}
  \end{center}
  \caption{Here we show the scatter of the maximum likelihood values
of the GWB amplitude and spectral index from the Monte-Carlo
simulations. From left to right the injected amplitudes are $1\times
10^{-15}$, $2\times 10^{-15}$, and $3\times 10^{-15}$ with spectral
index $13/3$ for a 10 pulsar PTA (top row) and 15 pulsar PTA (bottom
row). We can see that nearly all of these distributions display
minimal bias. } \label{fig:edf_scatter} \end{figure*}
In Figure \ref{fig:edf_scatter} we show the two-dimensional scatter
plot of the maximum likelihood parameters from our Monte-Carlo
simulations. It is clear that the bias in our two-dimensional
parameter space of interest is \emph{practically} very small. In fact
the means of the distributions for $A$ and $\gamma$ for the 10 pulsar
case are $(1.6,2.25,3.14)\times 10^{-15}$ and $(4.17,4.24,4.23)$,
respectively and for the 15 pulsar case we obtain
$(1.56,2.29,3.22)\times 10^{-15}$ and $(4.11,4.12,4.13)$,
respectively. In the first row of Figure \ref{fig:edf_scatter} we show
the 10 pulsar case with increasing GWB amplitude and the second row we
show the same for the 15 pulsar case. In the cases where there is a
bias present, the likelihood function prefers slightly lower spectral
indices and slightly larger amplitudes. However, from our experience
with the MDC this bias can also present itself by preferring a
slightly higher spectral index and lower amplitude. It should be noted
that even the smallest of the amplitudes tested here are near the
upper range of the expected level of the stochastic GWB \citep{s13}
and that the white noise rms of the pulsars is slightly unrealistic in
our current PTA regime. In fact we expect to have maybe five or six
pulsars that time at or below the 100 ns level while we have many
others that have much larger white noise rms. Thus we can conclude
that even though our likelihood is somewhat biased at larger
amplitudes (as is expected), for \emph{realistic} astrophysically
likely stochastic GWBs this method is effectively consistent and
unbiased. In fact, in terms of setting upper limits on the stochastic
GWB amplitude, this method is practically identical to using
the full likelihood, while much more computationally efficient.

\section{Discussion and Conclusions} \label{sec:conclusions} Here will
will briefly discuss future prospects of conducting a simultaneous
search for continuous GWs and the stochastic GWB. We will also compare
our work to other recent efforts to speed up PTA GW data analysis and
discuss the importance of our first-order likelihood method.

\subsection{Simultaneous Detection of Continuous GWs and a Stochastic
GWB}

One very important feature of the first order likelihood method is
that it can also be applied to searches for continuous GWs. This will
allow us to simultaneously search for a correlated stochastic
background and resolve individual sources that are bright enough to
stand out above such a background. In standard continuous GW searches
using PTAs \citep{bs12,esc12,pbs+12} the assumption is made that any
detectable single source will be bright enough such that the noise
(e.g stochastic GWB) can be approximated as a gaussian process that is
uncorrelated among pulsars. However, recent work \citep{rwh+12} has
shown that we are likely to see a few single sources per frequency bin
that will stand out from the typical isotropic stochastic background,
thus in order to resolve the weakest of these it is crucial to
simultaneously search for a correlated stochastic background as well
as the continuous source. We can then write down a combined likelihood
function assuming a deterministic source of functional form
$\mathbf{s}(\vec\lambda)$ \be
p(\mathbf{r}|\vec{\theta},\vec{\lambda})=\frac{1}{\sqrt{\det2\pi\boldsymbol{\Sigma}}}\exp\lp
-\frac{1}{2}(\mathbf{r}-\mathbf{s})^{T}
\boldsymbol{\Sigma}^{-1}(\mathbf{r}-\mathbf{s}) \rp,
\label{eq:singleFull} \ee where our noise (including the stochastic
background) parameters are $\vec\theta$ and our single source
parameters are $\vec\lambda$. Using our first order likelihood
approach we can approximate Eq. \ref{eq:singleFull} as
\begin{widetext} \be \begin{split}
\ln\,p(\mathbf{r}|\vec{\theta},\vec{\lambda})
&=\approx-\frac{1}{2}\left[ \Tr\ln \mathbf{P}
+(\mathbf{r}-\mathbf{s})^{T}\mathbf{P}^{-1}(\mathbf{r}-\mathbf{s})-(\mathbf{r}-\mathbf{s})^{T}\mathbf{P}^{-1}\mathbf{S}_{c}\mathbf{P}^{-1}(\mathbf{r}-\mathbf{s})
\right]\\ &=-\frac{1}{2}\sum_{\alpha=1}^M \left[\Tr\ln P_{\alpha}
+(r_{\alpha}-s_{\alpha})^TP_{\alpha}^{-1}(r_{\alpha}-s_{\alpha})
-\sum_{\beta\ne\alpha}^M(r_{\alpha}-s_{\alpha})^T
P_{\alpha}^{-1}S_{\alpha\beta}P_{\beta}^{-1}(r_{\beta}-s_{\beta})\right].
\end{split} \ee \end{widetext} As in the stochastic background case,
this again will speed up computations because we only have to invert
the \emph{individual} auto-covariance matrices as opposed to the
\emph{full} data covariance matrix. Although there have been proposed
methods to speed up the computation of the stochastic likelihood
function of Eq. \ref {eq:liker} \citep{vh12}, this is not applicable
to continuous sources because it relies on essentially applying a low
pass filter to the data. However, since we expect continuous sources
across the entire frequency band (with higher frequency sources
possibly standing out above the background) we must keep all frequency
information. Therefore our first order likelihood approximation is a
viable option when looking to significantly speed up computation time
while losing minimal information about potential GW signals.

As always, to claim a detection we must do some sort of model
comparison, be it a Neyman-Pearson test for Frequentist statistics or
an odds ratio or Bayes factor for Bayesian statistics. For example if
we want to assess the likelihood of that a continuous GW is in our
data we want to compute the following Bayes factor \be
\mathcal{B}=\frac{\mathcal{Z}_{\rm CW}}{\mathcal{Z}_{\rm
noise}}=\frac{\int\int
d\vec{\lambda}d\vec{\theta}p(\mathbf{r}|\vec\theta,\vec\lambda)p(\vec{\lambda})p(\vec{\theta})}{\int
d\vec{\theta}p(\mathbf{r}|\vec\theta)p(\vec{\theta})}, \ee where
$\mathcal{Z}_{\rm CW}$ and $\mathcal{Z}_{\rm noise}$ are the evidence
for the gravitational CW and noise models, respectively. However,
notice that $\vec\theta$ depends on our stochastic GWB parameters as
we treat all stochastic processes as ``noise'' in this analysis. If we
do not include the GWB parameters in the model then we could mistake a
low frequency GWB for a single continuous source, thus including the
GWB stochastic background in both models is crucial to detection and
eventually characterization of a single GW source. We should also
mention that the biases mentioned in section \ref{sec:edf} are not as
important if we simply wish to let the noise parameters vary along
with the single source parameters since these noise parameters will be
marginalized over in the end. An exploration of these combined
searches will be the subject of a future paper.

\subsection{Comparison with Other Work}

Recently there have been three studies devoted to making the analysis
of PTA data more computationally efficient. First, \citet[][hereafter
vH12]{vh12} have developed a method dubbed Acceleration By Compression
(ABC) to speed up this analysis. The main point of this work is to
write the data in a compressed basis, keeping the minimum number of
basis vectors to maximize the ability to characterize a correlated red
signal. This work also makes use of an interpolation scheme to compute
the covariance matrix which further improves the efficiency of the
algorithm at the cost of large memory usage. This method has proved to
be very efficient and accurate in setting upper limits on the
stochastic GWB and characterizing injected signals. However, since
this method relies on a reduced basis that essentially ``throws away''
high frequency information it is impossible to obtain a reliable Bayes
Factor when comparing models that allow for varying white noise
components. Since our first-order likelihood function makes use of all
the information in the data we can indeed compute reliable Bayes
factors and make confident statements about detection. We note however
that the first-order likelihood of this work and the ABC method of
vH12 are complementary. The two methods can in principle be combined
for even greater efficiency.

Most recently there have been two analyses of the IPTA MDC that aim to
make the PTA data analysis more efficient. First,
\citet{Lentati:2012xb} have developed a novel model-independent method
for the estimation of the spectral properties of an isotropic
stochastic GWB. This method uses a frequency domain approach and is
extremely efficient and results in computational speedups of two to
three orders of magnitude over the full likelihood implementation. It
has also been extensively tested on the MDC datasets and has proved to
be very accurate in characterizing the stochastic GWB. Our first order
likelihood method is indeed complementary to this work as it provides
a way to efficiently evaluate the likelihood function in a full time
domain analysis which will be vital for cross-checks for real-life
detection candidates.

Finally, \citet{Taylor:2012vx} have implemented the full VHML
likelihood function and have made it more efficient through the use of
optimized linear algebra libraries with multithreading and
parallelization resulting in significant speedups in the likelihood
evaluation. However, all of these methods could just as well be
applied to the first-order likelihood which would still be more
efficient than the full likelihood by a factor proportional to the
number of pulsars in the array.

This work and recent work have shown that there has indeed been
significant progress on making the likelihood evaluation more
efficient for pulsar timing arrays. All of these methods are
complementary and will provide important cross checks for future
stochastic GWB detection candidates.

\subsection{Summary}

In this paper we have introduced a novel way to speed up the
computation of the likelihood function for PTAs when searching for a
stochastic GWB. This was accomplished by expanding the likelihood
function to first order in the Hellings and Downs correlation
coefficients expected for a stochastic GWB leading to a computational
speedup on the order of the square of the number of pulsars in the
PTA. For typical PTAs this results in a speed-up of a few hundred to
about a thousand. We have briefly discussed the implementation of this
technique on the first IPTA Mock Data Challenge and showed that this
algorithm performs well in extracting the injected GWB parameters and
making a significant detection through various Bayes factors. Though
this is indeed an approximation to the full likelihood function we
have shown through extensive simulations that the bias introduced in
the estimation of GWB parameters is minimal and negligible in many
cases. This was accomplished through an analytical computation of the
expectation value of the maximum likelihood, direct comparisons of the
full and first-order likelihood functions on simulated data sets and
through a statistical Monte-Carlo approach based on the Empirical
Distribution Function. Although this work has focused solely on the
detection and characterization of a stochastic GWB, this likelihood
function can also be used to estimate the intrinsic red and white
noise parameters of individual pulsars simultaneously with the GWB
parameters.

\acknowledgements We would like to thank the members of the NANOGrav
detection working group for their comments and support, especially
Paul Demorest and Joe Romano. We would also like to thank Jolien
Creighton for useful conversations. This work was partially funded by
the NSF through CAREER award number 0955929, PIRE award number
0968126, and award number 0970074.

\appendix

\section{Relationship to VHML likelihood} \label{app:likelihood}
Making use of Eq. \ref{eq:resids}, the likelihood function for the
noise can be written as \be \begin{split}
p(\mathbf{n}|\vec\theta)=p(\mathbf{r}|\vec\theta,\delta\boldsymbol{\xi}_{\rm
best})=\frac{1}{\sqrt{\det(2 \pi \boldsymbol{\Sigma}_n)}}
\times\exp\lp-\frac{1}{2}(\mathbf{r}-\mathbf{M}\delta\boldsymbol{\xi}_{\rm
best})^T\boldsymbol{\Sigma}_n^{-1}(\mathbf{r}-\mathbf{M}\delta\boldsymbol{\xi}_{\rm
best})\rp. \end{split} \ee This can be thought of as a
\emph{conditional} pdf, where the values of
$\delta\boldsymbol{\xi}_{\rm best}$ are fixed. In \cite{vhl12} it was
shown that the marginalized likelihood can be written as \be
\begin{split} p(\mathbf{r}|\vec\theta)=\int
d\delta\boldsymbol{\xi}\,p(\mathbf{r}|\vec\theta,\delta\boldsymbol{\xi})
=\frac{\exp\left[ -\frac{1}{2}\mathbf{r}^{T}\mathbf{G}^{T}\lp
\mathbf{G}^{T}\boldsymbol{\Sigma}_{n}\mathbf{G}
\rp^{-1}\mathbf{G}^{T}\mathbf{r} \right]}{\sqrt{\det
2\pi\mathbf{G}^{T}\boldsymbol{\Sigma}_{n}\mathbf{G}}}, \end{split} \ee
where $\mathbf{G}$ is the matrix constructed from the final $(N-N_{\rm
fit})$ columns of the matrix $\mathbf{U}$ in the singular value
decomposition of the design matrix,
$\mathbf{M}=\mathbf{U}\mathbf{S}\mathbf{V}^{T}$.

We will now explore the $G$ matrix and the $R$ matrix obtained from
the marginalized and conditional pdfs, respectively. As mentioned
above, $R$ can be thought of as an oblique projection operator that
projects the pre-fit residuals into the post-fit residual space,
whereas $G^{T}$ can be thought of a projection operator that projects
our data onto the null space of $M$, that is, it projects the data
into a subspace orthogonal to the timing model fit. Since $R$ is not
generally symmetric and therefore is an oblique projection operator,
it does not have such a simple mathematical interpretation. However,
we can recast our problem in terms of ``weighted'' residuals then we
have the following transformations: $r\rightarrow Wr$, $M\rightarrow
WM$, and $R\rightarrow W^{-1}RW$, where $W$ is the weighting matrix
defined above. In this case minimizing the chi-squared becomes an
unweighted least squares problem and we obtain the exact same
estimates of $\delta\boldsymbol{\xi}_{\rm best}$ and likelihood
function as before. In this case $R$ is symmetric and can be thought
of as an orthogonal projection operator that projects our weighted
data onto the null space of the weighted timing model ($WM$). However,
in order to compute the likelihood we still have to invert the
covariance matrix $\Sigma_{r}=R\Sigma_{n}R^{T}$ which is singular. To
do this we rely on the pseudo-inverse. The pseudo-inverse of
$\Sigma_{r}$ is easiest defined in terms of its eigen-decomposition
$\Sigma_{r}=EDE^{T}$, with $E$ the matrix of eigenvectors of
$\Sigma_{r}$, and $D$ the diagonal matrix with $D_{ii}=\lambda_{i}$
the eigenvalues of $\Sigma_{r}$. It so happens that for a symmetric
positive semi-definite matrices like these, the eigen-decomposition is
also the singular value decomposition (SVD). The pseudo-inverse of
$\Sigma_{r}$ is then \be
\overline{\Sigma_{r}^{-1}}=E\overline{D^{-1}}D^{T}, \ee where the
overbar indicates that we are taking a pseudo-inverse and
$\overline{D^{-1}}_{ii}=1/\lambda_{i}$ for $\lambda>0$ and
$\overline{D^{-1}}_{ii}=0$ otherwise. Note that when all the error
bars are the same (i.e. $W=\sigma^{-1}\mathbb{I}$ with $\sigma$
constant), the matrix $G^{T}\Sigma_{n}G$ has the same eigenvalues as
the non-singular part of $R\Sigma_{n}R^{T}$ and we have \be
\overline{(R\Sigma_{n}R^{T})^{-1}}=G(G^{T}\Sigma_{n}G)^{-1}G^{T}. \ee
Thus we have obtained a very interesting result that in the case of
uniform uncertainties, the conditional pdf making use of a
pseudo-inverse is equivalent to the marginalized pdf making use of the
projection matrix $G^{T}$. However, in general this is not true and
the two methods are indeed different. Although, in many cases the
uncertainties are similar on a majority of the TOAs, thus the two
methods will not differ much in practice.

\bibliographystyle{apj} \bibliography{apjjabb,bib}

\end{document}